\documentclass[10pt,conference,final,twocolumn]{IEEEtran}
\usepackage[
    bottom=1.02in,
    top=0.7in,        
    left=0.65in,
    right=0.65in      
]{geometry}
\usepackage[T1]{fontenc}
\usepackage{cite}
\usepackage{amsthm}
\usepackage{amssymb}
\usepackage{amsfonts}
\usepackage{fancyhdr}  %
\usepackage{extarrows}
\usepackage{algorithm}
\usepackage{multirow,tabularx}
\usepackage{mathtools}
\usepackage{xcolor}
\usepackage[english]{babel}
\usepackage{caption}
\usepackage{bm}
\usepackage{algorithmic}
\usepackage{amsmath}
\usepackage{subfigure}
\usepackage{makecell}
\usepackage{colortbl}
\usepackage{booktabs}
\usepackage{graphicx}
\usepackage{diagbox}
\usepackage{cancel}
\usepackage{tablefootnote}
\usepackage{makecell}

\usepackage[bookmarks=false]{hyperref} 
\usepackage{cleveref} 
\usepackage{setspace}
\hypersetup{colorlinks = true, 
	    linkcolor = black, 
	    urlcolor = black,
            citecolor = black} 

\newtheorem{lemma}{Lemma}

\begin{document}
\bstctlcite{IEEEexample:BSTcontrol}  
\title{
Communication-Computation Pipeline Parallel Split Learning over Wireless Edge Networks
}
\author{
Chenyu~Liu,
Zhaoyang~Zhang\textsuperscript{$\dagger$}, 
Zirui~Chen, 
and Zhaohui~Yang \\
\IEEEauthorblockA{
    College of Information Science and Electronic Engineering, Zhejiang University, Hangzhou 310027, China\\
    Zhejiang Provincial Lab of Multi-Modal Commun. Networks \& Intelligent Info. Proc., Hangzhou 310027, China\\
    E-mails: \{chenyuliu, ning\_ming\textsuperscript{$\dagger$}, ziruichen, yang\_zhaohui\}@zju.edu.cn
    \vspace{-0.3cm}
}
}

\maketitle 
\begin{abstract}
Split learning (SL) offloads main computing tasks from multiple resource-constrained user equippments (UEs) to the base station (BS), while preserving local data privacy. However, its computation and communication processes remain sequential, resulting in limited system efficiency. To overcome this limitation, this paper applies pipeline parallelism (PP) of distributed training to SL in wireless networks, proposing the so-called communication-computation pipeline parallel split learning (C$^2$P$^2$SL). By considering the communicating and computing processes of UEs and BS as an overall pipeline, C$^2$P$^2$SL achieves pipeline parallelization among different micro-batches which are split from each batch of data samples. The overlap of communication and computation in this way significantly reduces the total training time. Given that training efficiency is affected by position of cutting layer and heterogeneity of the UEs, we formulate a joint optimization problem of task split and resource allocation, and design a solution based on alternating optimization. Experimental results demonstrate that C$^2$P$^2$SL significantly reduces system training time by over 38\% while maintaining convergence accuracy under different communication conditions.

\end{abstract}
\IEEEpeerreviewmaketitle
\vspace{-0.2cm}
\section{Introduction}\label{section1}
\vspace{-0.1cm}
In wireless edge networks, user equipments (UEs) are typically resource-constrained private devices, such as smartphones, laptops, and drones. Insufficient computation and storage capabilities make it difficult to meet the requirements for complex model training\cite{BAIM}. It is necessary to offload computational tasks from multiple UEs to base station (BS) with adequate resources via uplink transmission (UT) and downlink transmission (DT)\cite{WSL, USL}. However, for artificial intelligence (AI) driven application scenarios that involve personal information, such as image recognition and anomaly detection monitoring, privacy concerns associated with local data hinder direct uploading to the BS, thereby introducing challenges in distributed training between UEs and the BS.

Split learning (SL) is a promising approach that scatters deep neural networks (DNNs) to clients and server for collaborative training. Specifically, the initialized model is split into two parts: the first few layers are offloaded to the clients, while the main body of the model is retained on the server side\cite{SL}. SL uploads client-side model's outputs to the server, thereby keeping raw data samples localized and preventing privacy leakage. But sequential training across different clients can introduce excessive computational latency. To mitigate this, the works in\cite{SF} and \cite{SFL} propose split federated learning (SFL), which parallelizes client-side model training while periodically averaging model parameters, at the cost of potential privacy leakage risks and increased communication overhead.

The non-negligible communication latency is mainly because data transmission between the BS and UEs over the wireless network relies on uplink and downlink channels. To address this, \cite{PSL} proposes parallel split learning (PSL), eliminating the need for synchronization of client-side models based on SFL. Besides, the authors in\cite{EPSL} develop efficient parallel split learning (EPSL), which reduces the dimension of backward propagation activations through gradient aggregation of the last layer, thereby decreasing server-side training and communication latency. However, these methods mainly focus on reducing the volume of data transmission during the training process. For large models with high-dimensional intermediate outputs or scenarios with poor channel conditions\cite{WBAIM}, excessive communication time still dominates the training process, resulting in poor training efficiency. 

Meanwhile, among numerous parallelization strategies for distributed training with multiple graphics processing units (GPUs), pipeline parallelism (PP) operates similarly to SL, enhancing training efficiency through pipelined task execution \cite{GPipe, FTpipehd}. Specifically, the model is partitioned into sub-models distributed across distinct GPUs, with sequential training stages forming an execution pipeline. Unlike conventional training where each iteration processes one batch, PP further partitions batches into multiple micro-batches. Each GPU propagates the results to the subsequent stage immediately after completing its sub-model computation on the current micro-batch, while concurrently processing intermediate outputs from the preceding stage for subsequent micro-batches. This mechanism achieves fine-grained parallelism by overlapping computation of micro-batches.

Inspired by the similar mechanism, in this paper, we propose the communication-computation pipeline parallel split learning (C$^2$P$^2$SL), which integrates PP into PSL frameworks. In C$^2$P$^2$SL, the computation pipeline of UEs and the BS is combined with the communication pipeline consisting of uplink and downlink time slots. By splitting input batches into multiple micro-batches, C$^2$P$^2$SL achieves parallelization between computation and communication pipelines of different micro-batches. Moreover, considering that the efficiency of whole model training is determined by task partition and heterogeneous capabilities of the UEs, we formulate a joint optimization problem to minimize the bubble rate, which is the ratio of idle time to total time. To solve this problem, we adopt an alternating optimization (AO) based split and allocation method. The main contributions of this paper are summarized as follows:
 
\begin{itemize}
    \item We propose C$^2$P$^2$SL over wireless edge networks, a highly parallelized SL framework that schedules the computation and communication tasks of UEs and BS in parallel.
    \item We design an AO-based method to jointly optimize model and batch splitting together with batch size and time slots allocation to improve the training efficiency.
    \item We conduct experiments under various UE numbers to demonstrate the effectiveness of C$^2$P$^2$SL. Experimental results show significant improvements in training efficiency and robustness under varying bandwidths.
\end{itemize}

\section{System Model and C$^2$P$^2$SL Framework}\label{system model}
In this section, we introduce the computation and communication models of C$^2$P$^2$SL, as well as the training workflow.
\vspace{-0.5cm}
\subsection{Computation Model}\label{section2.1}
\begin{figure}[h!]
    \centering
    \includegraphics[width=0.46\textwidth]{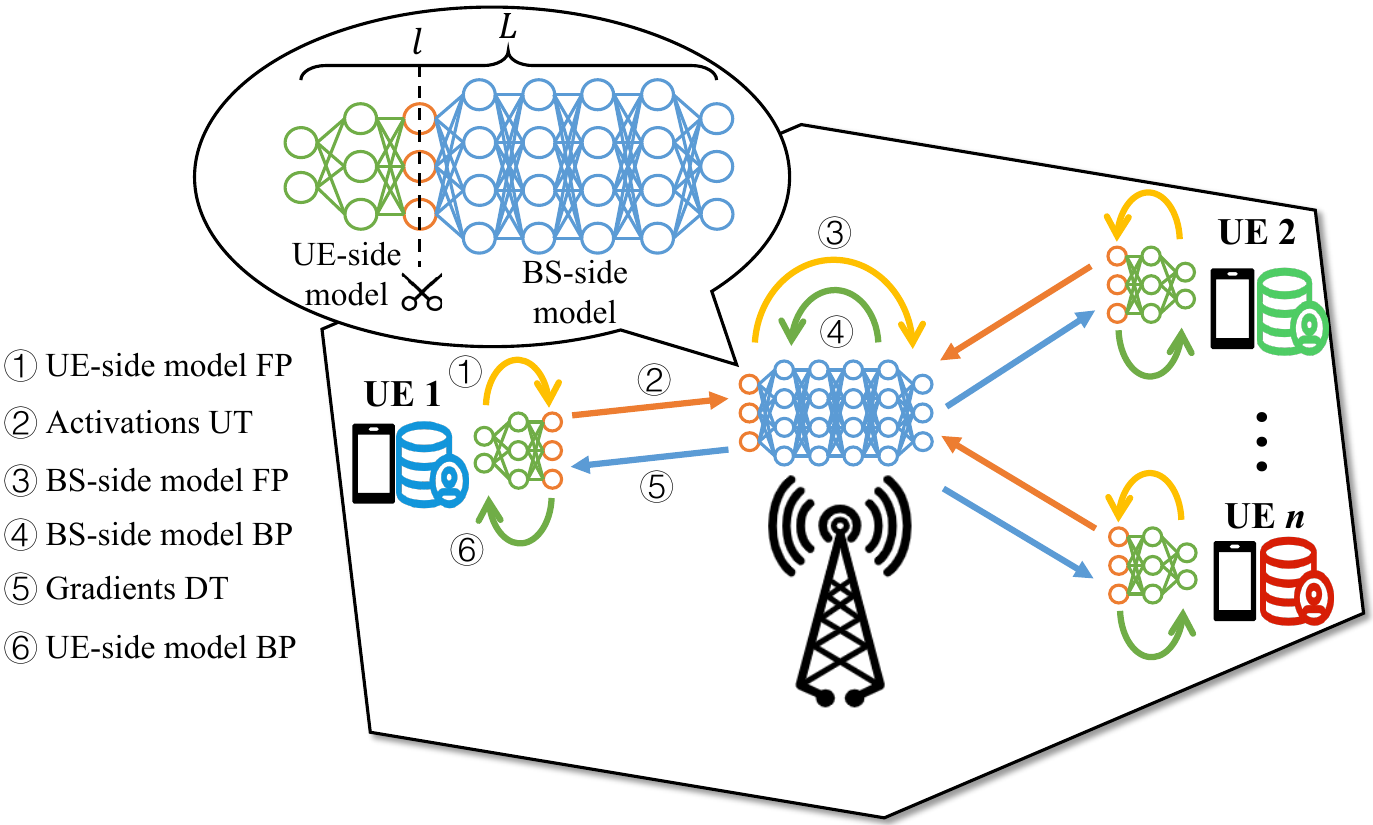}
    \caption{The proposed C$^2$P$^2$SL over edge networks.}
    \label{framework}
\end{figure}

As illustrated in Fig.~\ref{framework}, we consider a cellular network comprising a central BS and \(n\) UEs. Each UE has weak computational capability to execute forward propagation (FP) and backward propagation (BP) of tiny neural network, along with locally stored private data samples. In contrast, by integrating AI processing units, the BS has sufficient computing resources to support training of larger neural networks. Let \(N=\{1,2,\ldots,n\}\) denote the set of UEs.
\vspace{-0.3cm}
\vspace{-0.3cm}

The target model has a total of \(L\) layers, which are partitioned into UE-side model and BS-side model before training. The indice of cut layer is denoted by $l$, \(l\in\{1,2,\ldots,L\}\). In this way BS retains the submodel containing the majority of parameters, while broadcasting the smaller submodel to individual UEs. Note that for the same network layer, the activation gradients generated by BP have the same dimension as the activation outputs generated by FP. We indicate the data size of cutting layer's output as \(s_l\).

When inputting a single data sample into the neural network, the $j$-th layer's computational workload for FP and BP are respectively denoted by \(c_{j}^{F}\) and \(c_{j}^{B}\). During one training round, the batch size of the $i$-th UE's input data is denoted by \(b_i\), and the total batch size of all UEs is given as
\begin{equation}b=\sum_{i=1}^{n}b_i.\end{equation}

Given that device performance remains stable over short periods, we consider the computational capacities of the $i$-th UE and the BS to be constant during training a batch of data samples, represented as
\begin{equation}f_i=K_UF_i,\end{equation}
\begin{equation}f_B=K_BF_B,\end{equation}
where \(F_i\) and \(F_B\) represent the clock frequencies of the $i$-th UE and BS, respectively. \(K_U\) and \(K_B\) denote the corresponding floating point operations per clock cycle of UE and BS.
\vspace{-0.1cm}
\subsection{Communication Model}\label{section2.2}
We consider a time division multiple access (TDMA) communication system in which all UEs share the total system bandwidth $B$. TDMA partitions time into periodic time frames of length $T$, with each frame subdivided into multiple time slots. Each slot is utilized for transmitting one UE's data and the length of time slot allocated to UE $i$ is denoted by \(\tau_i\). Multiple access is achieved by allocating time slots within each frame to individual UEs. Thus, the relationship between time frame and slots is expressed as
\begin{equation}\sum_{i=1}^n\tau_i \le T.\end{equation}

Noting that the channel conditions remain approximately stable within a short time slot, the communication capability of each UE is considered constant during a communication round. Consequently, based on Shannon's theorem under additive white Gaussian noise (AWGN) channel, the achievable uplink and downlink rates of $i$-th UE and BS can be respectively expressed as
\begin{equation}r_u^i=B\log_2\left(1+\frac{Gp_i h(d_i,f)}{BN_0}\right),\end{equation}
\begin{equation}r_d^i=B\log_2\left(1+\frac{Gp_B h(d_i,f)}{BN_0}\right),\end{equation}
where \(p_i\) denotes the uplink transmit power of UE and \(p_B\) represents the downlink transmit power of the BS. \(G\) stands for combined antenna gain of transmitter and receiver. \(h(d_i,f)\) is the path loss of UE \(i\), modeled as a linear function of the UE-BS distance \(d_i\) and the system frequency \(f\). \(N_0\) denotes the power spectral density (PSD) of noise.
\vspace{-0.1cm}
\subsection{C$^2$P$^2$SL Workflow}\label{section2.3}
\begin{figure}[htbp]
    \centering
    \includegraphics[width=0.48\textwidth]{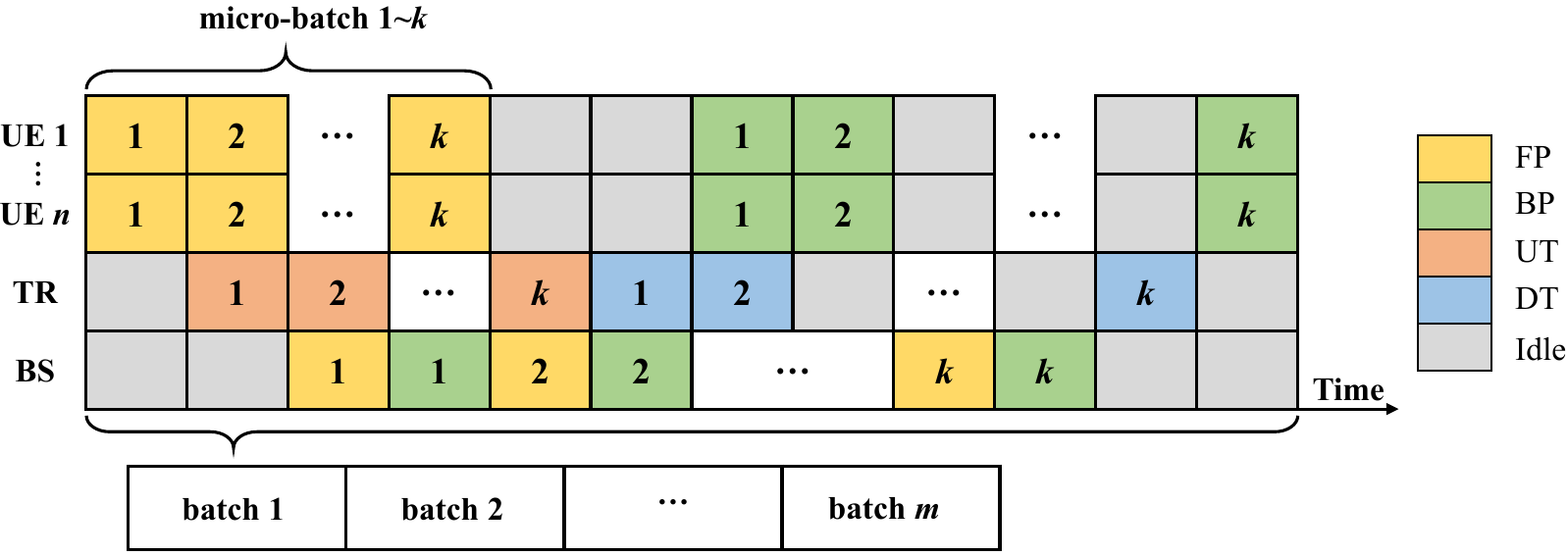}
    \caption{The training workflow of C$^2$P$^2$SL.}
    \vspace{-0.65cm}
    \label{schedule}
\end{figure}

After model initialization and distribution, the training process is shown in Fig.~\ref{schedule}, where the vertical axis sequentially represents the tasks of each UE, communication link, and BS while blocks of the horizontal axis represent the time of each task. For convenience, we describe training iteration of a single batch, which is divided into $k$ micro-batches.

\subsubsection{Forward Stage}\label{section2.3.1}
Since BS broadcasts the same UE-side model to all UEs, the parameters are identical before training begins. After loading the local data, UEs sequentially input the split $k$ micro-batches into the UE-side models for FP. Constrained by inter-UE heterogeneity, different computational capabilities result in varying completion times across UEs, thereby incurring additional latency. The calculation time for a micro-batch of UE $i$ is denoted by 
\begin{equation}t_i^F=\frac{b_i\sum_{j=1}^lc_j^F}{kf_i}.\end{equation}

Once a micro-batch's FP is completed, UEs upload the activation outputs and the corresponding labels to BS, while continuing with the next FP. A UT block is used to represent the combination of UT time slots of all UEs in Fig. \ref{schedule}. The UT time of UE $i$ for each micro-batch is calculated as
\begin{equation}
t_i^U=\frac{b_i(s_l+s_0)T}{kr_u^i\tau_i},
\end{equation}
where \(s_0\) denotes the label size of a data sample. After receiving activation outputs from all UEs of micro-batch with the same index, the BS aggregates them and executes FP of BS-side model, while receiving the next uplink data. After that, the BS computes the loss by the output of BS-side model and corresponding labels. The computation time is given by
\begin{equation}t_b^F=\frac{\sum_{i=1}^nb_i\sum_{j=l+1}^Lc_j^F}{kf_B}.\end{equation}

We define the forward stage as spanning from the start of FP for the first micro-batch to the completion of UT for the final micro-batch. Therefore, the forward process composed of FP and UT tasks is executed sequentially for a single micro-batch, while different tasks of distinct micro-batches run in parallel.

\subsubsection{Backward Stage}\label{section2.3.2}
The computation tasks on the BS side follow the one forward pass followed by one backward pass (1F1B) principle: After completing the BS-side model FP of a micro-batch, BS immediately executes the BS-side model BP based on the loss. The computation time of BP for a micro-batch is described as
\begin{equation}t_b^B=\frac{\sum_{i=1}^nb_i\sum_{j=l+1}^Lc_j^B}{kf_B}.\end{equation}

After completing the BP of the current micro-batch, BS scatters and distributes the activation gradients obtained by BS-side model BP to UEs based on the FP outputs, while continuing to execute the next FP and BP. The DT time of UE $i$ for each micro-batch can be given as
\begin{equation}
t_i^D=\frac{b_is_lT}{kr_d^i\tau_i}.
\end{equation}

As illustrated in Fig.~\ref{schedule}, subsequent UT processes may still be ongoing at this stage. This potentially cause conflicts between UT and DT processes because of the same frequency band. Given that the total completion time is only constrained by BP of the final micro-batch in UE-side model, we prioritize UT to ensure orderly training progression. That is, the DT process at the BS needs to commence after all micro-batches have completed UT, which induces partial latency only under poor communication conditions.

Receiving the gradients from BS, UEs have to complete the BP process of UE-side model so that they could update the parameters of their models later. The BP time of UE $i$ is denoted as 
\begin{equation}t_i^B=\frac{b_i\sum_{j=1}^lc_j^B}{kf_i}.\end{equation}

Upon termination of backward stage, UEs immediately execute updates to UE-side models and BS concurrently updates the parameters after the last micro-batch's BP of BS-side model. Although each micro-batch is processed independently, the accumulated gradients are mathematical equivalence to full batch computation, thereby ensuring normal model updates throughout the system.
\vspace{-0.1cm}
\section{Optimization for Split and Allocation in C$^2$P$^2$SL}\label{section3}
In this section, we formulate system constraints as a joint optimization problem and design an AO-based solution.
\vspace{-0.1cm}
\subsection{Constraints Formulation}\label{section3.1}
In C$^2$P$^2$SL, the cut position of the neural network model directly determines computational tasks of UEs and BS, as well as data transmitted over uplink and downlink channels, thereby affecting computation and communication durations. Furthermore, the micro-batch size per training round influences the degree of parallelism among tasks. Computation and communication times vary across UEs due to heterogeneous workloads and device capabilities. Subsequently, quantitative analysis and formulation among these constraints will be conducted.

\subsubsection{Storage Limitation}\label{section3.1.1}
Due to constrained storage resources of personal devices, UEs cannot accommodate large model weight parameters, optimizer states, and intermediate activations. Since both computational load and memory storage increase monotonically with the depth of neural network, we represent the storage boundary by using the computational load $c_{i}$ at the time of reaching the maximum memory capacity of UE $i$. It depends mainly on device storage capabilities and model architecture. This implies that model split must satisfy the constrained condition as

\begin{equation}
    \sum_{j=1}^l(c_j^F+c_j^B)b_i \le c_i, \forall i \in N.
\end{equation}

\subsubsection{Idle Time}\label{section3.1.2}
As discussed in the previous section, durations of computational and communication tasks exhibit significant variation. Given the sequentiality of computation and communication tasks, subsequent tasks will be affected by longer preceding tasks. When preceding tasks experience prolonged execution times, situations occur where a downstream processor completes its current task while the upstream processor has not yet finalized the next task. This mismatch generates additional idle time during pipeline progression. To ensure training efficiency and continuity of BS which is the main work units, such idle periods must be minimized, requiring the forward stage to satisfy the constraints as
\vspace{-0.05cm}
\begin{equation}
    \max\{\max_it_i^F,\max_it_i^U\}\le t_b^F+t_b^B.\label{FS}
\end{equation}
\vspace{-0.45cm}

For the backward stage, we have mentioned that the end time of training a batch is determined by BP of the final micro-batch. Therefore, it is necessary to ensure that there is no additional waiting time in the computational and communication process of the final micro-batch, which requires the backward stage to satisfy
\vspace{-0.05cm}
\begin{equation}
    (k-1)(\max_it_i^U+\max_it_i^D)\le k(t_b^F+t_b^B).\label{BS}
\end{equation}
\vspace{-0.45cm}

\subsubsection{Bubble Rate}\label{section3.1.3}
To reasonably measure the training efficiency, we use the bubble ratio as a metric to quantify efficiency, defined as the proportion of idle time to total training time. It can be expressed as
\vspace{-0.05cm}
\begin{equation}
    BR=\frac{t_{idle}}{t_{idle}+t_{work}}.
\end{equation}
\vspace{-0.3cm}

It is easy to see that a lower bubble ratio indicates higher training efficiency. Given that the computational capacity of BS significantly exceeds that of UEs, we adopt the BS-side bubble ratio as the representative efficiency metric. Under the constraints of \eqref{FS} and \eqref{BS}, the idle time of BS can be calculated respectively, as
\vspace{-0.05cm}
\begin{equation}
    t_{idle}=\max_i(t_i^F+t_i^U)+\max_i(t_i^D+t_i^B).
\end{equation}
\vspace{-0.2cm}
And the work time of BS is formulated as
\vspace{0.1cm}
\begin{equation}
    t_{work}=k(t_b^F+t_b^B).
\end{equation}
\vspace{-0.45cm}

Therefore, under local UE resource constraints while ensuring data privacy guarantees, we optimize the model partition position $l$, micro-batch division count $k$, UE batch size set $\boldsymbol{b}$, and allocated time slot set $\boldsymbol{\tau}$ to minimize bubble ratio during training. Based on computation-communication time analysis, we formulate a joint optimization problem of split and allocation, which can be modeled as
\vspace{-0.05cm}
\begin{align} \label{P1}
P1: \quad & \min_{k,l,\boldsymbol{b},\boldsymbol{\tau}} BR(k,l,\boldsymbol{b}, \boldsymbol{\tau})\\
\text{s.t.} \quad & \text{C1: }1 \le l \le L-1,\nonumber\\
& \text{C2: }\sum_{j=1}^l(c_j^F+c_j^B)b_i \le c_i, \forall i \in N,\nonumber\\
& \text{C3: }\max\{\max_it_i^F,\max_it_i^U\}\le t_b^F+t_b^B, \nonumber\\
& \text{C4: }(k-1)(\max_it_i^U+\max_it_i^D)\le k(t_b^F+t_b^B), \nonumber\\
& \text{C5: }\sum_{i=1}^nb_i=b, b_i\ge0,\nonumber\\
& \text{C6: }\sum_{i=1}^n\tau_i\le T, \tau_i\ge0.\nonumber
\end{align}
\vspace{-0.5cm}

\vspace{0.1cm}
\subsection{Joint Optimization}\label{section3.2}
Since the decision variables $l$, $k$, $b_i$ are integers, with nonlinear and coupled objective functions and constraints, $P1$ constitutes a mixed-integer nonlinear programming (MINLP) problem. Non-convexity of the objective function precludes a direct optimal solution, while the constraints have joint restrictions on the variables. To address this, we adopt an AO-based method, breaking $P1$ into three subproblems\cite{AO}.

\subsubsection{Model Layer and Micro-batch Split}\label{section3.2.1}
First we fix $\boldsymbol{b},\boldsymbol{\tau}$, P1 is transformed into an integer nonlinear programming (INLP) problem concerning the partitioning variables $l$ and $k$. Given that the feasible range of $l$ depends on the total layers $L$, we enumerate all possible partitioning points by the constraints of C1,C2,C3. For each available $l$, the corresponding model parameters $c_i^F, c_i^B, s_l$ can be regarded as constants, thus obtaining a problem of maximizing $k$. By calculating C4, we have lemma \ref{lemma} to determine the value of $k$.

\begin{lemma}\label{lemma}
    For every fixed $(l,\boldsymbol{b},\boldsymbol{\tau})$, the optimal micro-batch number
    $k=\left \lfloor\frac{1}{1-\eta}\right \rfloor,  \eta=\mathop{\max}\limits_i\frac{\tau_ib\sum_{j=l+1}^L(c_j^F+c_j^B)}{b_iTf_B\left((s_l+s_0)/r_u^i+s_l/r_d^i\right)}$.
\end{lemma}
After enumerating all possible $l$ with the corresponding $k$ following lemma \ref{lemma}, we compare and select the optimal $(l,k)$ to minimize the bubble rate.

\subsubsection{Batch Size partition}\label{section3.2.2}
By fixing $(l,k)$ and $\boldsymbol{\tau}$, we focus on the subproblem addressing batch size partition. Thus, $P1$ can be reduced to $P2$ as shown below. 
\begin{align} \label{P2}
P2: \quad & \min_{\boldsymbol{b}} \{\max_i(t_i^F+t_i^U)+\max_i(t_i^D+t_i^B)\}\\
\text{s.t.} \quad & \text{C2, C3, C4, C5.}\nonumber
\end{align}

$P2$ is a nonlinear optimization problem with a nonsmooth function. Since the maximum term in the objective function and constraints can be transformed into linear constraints by introducing auxiliary variables $t_1=\mathop{\max}\limits_i(t_i^F+t_i^U)$, $t_2=\mathop{\max}\limits_i(t_i^D+t_i^B)$, $t_3=\mathop{\max}\limits_i(t_i^U)$, and $t_4=\mathop{\max}\limits_i(t_i^D)$, the problem can be equivalently transformed into $P3$. As a mixed-integer linear programming (MILP) problem, $P3$ is easy to solve by using available toolkits such as CVXPY \cite{CVX}.
\vspace{-3pt} 
\begin{align}\label{P3}
P3: \quad & \min_{\boldsymbol{b},t_1,t_2,t_3,t_4}(t_1+t_2)\\
\text{s.t.} \quad & \text{C2, C3, C5},\nonumber\\
& \widetilde{\text{C}}4: (k-1)(t_3+t_4)\le k(t_b^F+t_b^B),\nonumber\\
& \text{C7: }t_i^F+t_i^U\le t_1, \forall i \in N,\nonumber\\
& \text{C8: }t_i^D+t_i^B\le t_2, \forall i \in N,\nonumber\\
& \text{C9: }t_i^U\le t_3, \forall i \in N,\nonumber\\
& \text{C10: }t_i^D\le t_4, \forall i \in N.\nonumber
\end{align}

\subsubsection{Time Slot Allocation}\label{section3.3.3}
By fixing $(l,k,\boldsymbol{b})$, we turn $P1$ into a nonlinear and nonsmooth problem. Similarly, we have $P4$ by introducing the same auxiliary variables as above.
\begin{align}\label{P4}
P4: \quad & \min_{\boldsymbol{\tau},t_1,t_2,t_3,t_4}(t_1+t_2)\\
\text{s.t.} \quad & \widetilde{\text{C}}4,\text{C6, C7, C8, C9, C10}.\nonumber\\
& \widetilde{\text{C}}3: t_i^U\le t_b^F+t_b^B, \forall i \in N.\nonumber
\end{align}

Given that the constraints of $\widetilde{\text{C}}$3,C7,C8,C9,C10 on $\tau_i$ are all lower bounds, we use the function $g_i(t_1,t_2,t_3,t_4)$ to represent the joint constraint of $\tau_i$. Therefore, by representing $\tau_i$ with other variables, $P4$ can be converted to $P5$, which is a convex optimization problem that can be solved using CVXPY\cite{CVX}.
\vspace{-0.1cm}
\begin{align}\label{P5}
P5: \quad & \min_{t_1,t_2,t_3,t_4}(t_1+t_2)\\\
\text{s.t.} \quad & \sum_{i=1}^ng_i(t_1, t_2, t_3, t_4)\le T,\nonumber\\
& t_3+t_4\le \frac{k}{k-1}(t_b^F+t_b^B),\nonumber\\
& t_1\ge t_i^F, t_2\ge t_i^B, \forall i \in N,\nonumber\\
& t_3\ge0, t_4\ge0.\nonumber
\end{align}
\vspace{-0.5cm}

As mentioned above, the original problem $P1$ is decomposed into three tractable subproblems. Then, we design a split and allocation algorithm based on AO, which is described in Algorithm \ref{algo}. The complexity of Algorithm \ref{algo} depends mainly on P3. Although MILP problems are NP-hard with potentially exponential time complexity growth as the number of UEs increases, the constraint C5 enables simplification via branch and bound. Specifically, we solve the corresponding LP relaxation by relaxing integer constraints and then compare flooring and ceiling solutions of the results. In this way, the total complexity of the algorithm is $O(log(\frac{1}{\epsilon})\cdot(L+n^3))$.

\vspace{-0.2cm}
\begin{algorithm}[htbp]
\caption{Split and Allocation Algorithm Based on AO} \label{algo}
\begin{algorithmic}[1]
\REQUIRE Convergence tolerance $\epsilon$, iteration index $m=0$.
\ENSURE $k^*, l^*, \boldsymbol{b}^*,\boldsymbol{\tau}^*$. 
\STATE Initialization: $k^{(0)}, l^{(0)}, \boldsymbol{b}^{(0)}, \boldsymbol{\tau}^{(0)}$
\REPEAT
    \STATE $m \gets m+1, BR \gets \infty$
    \FOR{$l=1$ to $L$}
        \STATE Obtain $k$ by lemma \ref{lemma}.
        \IF {$BR(l,k) < BR_{\min}$}
            \STATE$BR_{\min} \gets BR(l,k), l^{(m)} \gets l, k^{(m)} \gets k$.
        \ENDIF
    \ENDFOR
    \STATE Update $\boldsymbol{b}^{(m)}$ by solving MILP problem $P3$ (\ref{P3}).
    \STATE Update $\boldsymbol{\tau}^{(m)}$ by solving convex problem $P5$ (\ref{P5}).
\UNTIL{$|BR(k^{(m)},l^{(m)}, \boldsymbol{b}^{(m)}, \boldsymbol{\tau}^{(m)})-BR(k^{(m-1)},l^{(m-1)},$ $\boldsymbol{b}^{(m-1)},\boldsymbol{\tau}^{(m-1)})| \leq\epsilon$.}
\end{algorithmic}
\end{algorithm}
\vspace{-0.4cm}

\section{Performance Evaluation}\label{section4}
In this section, we perform experiments under different situations to evaluate the performance of C$^2$P$^2$SL.
\vspace{-0.2cm}
\subsection{Experiment Settings}\label{section4.1}
We consider a cellular network with a 500\,\text{m} radius where the BS is centrally located and UEs are randomly distributed. Each UE's clock frequency is randomly selected within [0.5, 1.5]\,\text{Gcycles/s}, while the transmit powers follow random distributions within [13, 23]\,\text{dBm}. These parameters reflect performance heterogeneity among UEs in wireless networks. And we use $h(d_i,f)=28.0+22\log_{10}d_i+20\log_{10}f$\,\text{dB} to denote the path loss function of UE $i$. The remaining parameters are specified in Table \ref{system}.
\begin{table}[htbp]
\renewcommand{\arraystretch}{1.3}
\centering 
\footnotesize
\caption{System parameter settings}
\vspace{-0.2cm}
\label{system}
\begin{tabular}{c|c|c|c}
\bottomrule[0.5mm]
\textbf{Parameter} & \textbf{Value} & \textbf{Parameter} & \textbf{Value}\\ \hline
$f$   & 3.5\,GHz & $B$ & 100\,MHz\\
$d_i$ & [100, 500]\,m & $T$ & 10\,ms\\
$p_i$ & [13, 23]\,dBm & $p_B$ & 46\,dBm \\ 
$G$   & 10 & $N_0$ & -174\,dBm/Hz\\
$F_i$ & [1, 2]\,Gcycle/s & $F_B$ & 80\,Gcycle/s\\ 
$K_U$ & 16\,FLOPs/cycle & $K_B$ & 32\,FLOPs/cycle\\ 
$b$   & 512& $c_i$ & [1, 2]\,GFLOPs\\ \hline
\toprule[0.5mm]
\end{tabular}
\vspace{-0.6cm}
\end{table}

The ResNet-18\cite{RESNET} network which has 18 layers and CIFAR-10 dataset\cite{CIFAR} are adopted for training. To accommodate the 32×32 dimensions of CIFAR-10, we modify specific model parameters to enable proper training. Computational workload and communication volume per layer inputting a single sample are detailed in Table \ref{model}, where blocks 1$\sim$4 contain multiple convolutional layers. To ensure training continuity, the 50,000 CIFAR-10 training samples are distributed to UEs proportionally to their batch sizes before training begins while the test dataset comprises 10,000 samples.
\vspace{-0.1cm}
\begin{table}[htbp]
\renewcommand{\arraystretch}{1.3}
  \centering
  \footnotesize
  \caption{Model parameters of ResNet-18}
  \vspace{-0.2cm}
  \label{model}
  \begin{tabular}{c|c|c|c}
    \bottomrule[0.5mm]
    \textbf{Layer}     & \textbf{Params (M)} & \textbf{FLOPs (MFLOP)} & \textbf{Traffic (MB)} \\ \hline
    Conv1   & 0.002      & 3.802         & 0.250        \\ 
    Block1  & 0.148      & 303.0        & 0.250        \\ 
    Block2  & 0.526      & 269.1        & 0.125        \\ 
    Block3  & 2.100      & 268.8        & 0.063        \\ 
    Block4  & 8.394      & 268.6        & 0.031        \\ 
    Avgpool+FC & 0.005   & 0.026         & 3.81E-05     \\ \hline
    \toprule[0.5mm]
  \end{tabular}
\end{table}
\vspace{-0.2cm}

To validate the effectiveness of the proposed scheme, we compare it against three representative benchmarks:
\begin{itemize}
    \item SL\cite{SL}: The BS trains individually with UEs through model partitioning, requiring each UE to sequentially train UE-side model by local data.
    \item PSL\cite{PSL}: Shared BS-side model is trained using aggregated data from all UEs, while UE-side models are trained using individual UE data in parallel.
    \item EPSL\cite{EPSL}: Based on PSL, the activation gradient dimension of BP is reduced by the last layer of gradient aggregation of the last layer.
\end{itemize}

\subsection{Experimental Results}\label{section4.2}
\vspace{-0.4cm}
\begin{figure}[ht]
    \centering 
    \includegraphics[width=0.3\textwidth]{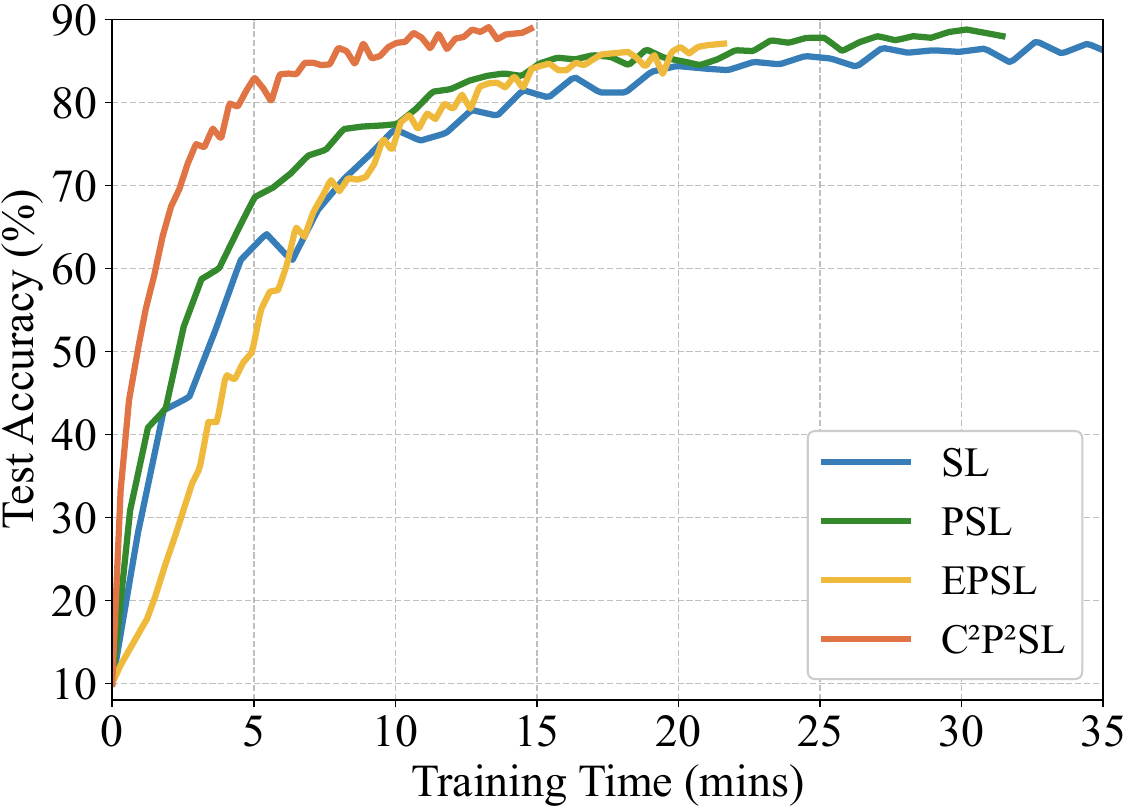}
    \vspace{-0.15cm}
    \caption{Test accuracy of various schemes with n=8.}
    \vspace{-0.3cm}
    \label{accuracy}
\end{figure}
Fig.~\ref{accuracy} shows the testing accuracy during the training process of various schemes when the number of UEs is 8. It can be seen that our C$^2$P$^2$SL scheme maintains convergence accuracy close to PSL and SL, slightly higher than EPSL, which aggregates the gradients of different UEs during BP. As stated in Section \ref{system model}, C$^2$P$^2$SL does not change the FP and BP process of a single micro-batch. For benchmark schemes, training convergence time of sequential SL includes computation time of UE and BS, as well as communication time between. Due to the overlap of computation time and communication time, we significantly reduce the convergence time.

\vspace{-0.25cm}
\begin{figure}[ht]
    \centering 
    \hspace{0.6cm}
    \includegraphics[width=0.36\textwidth]{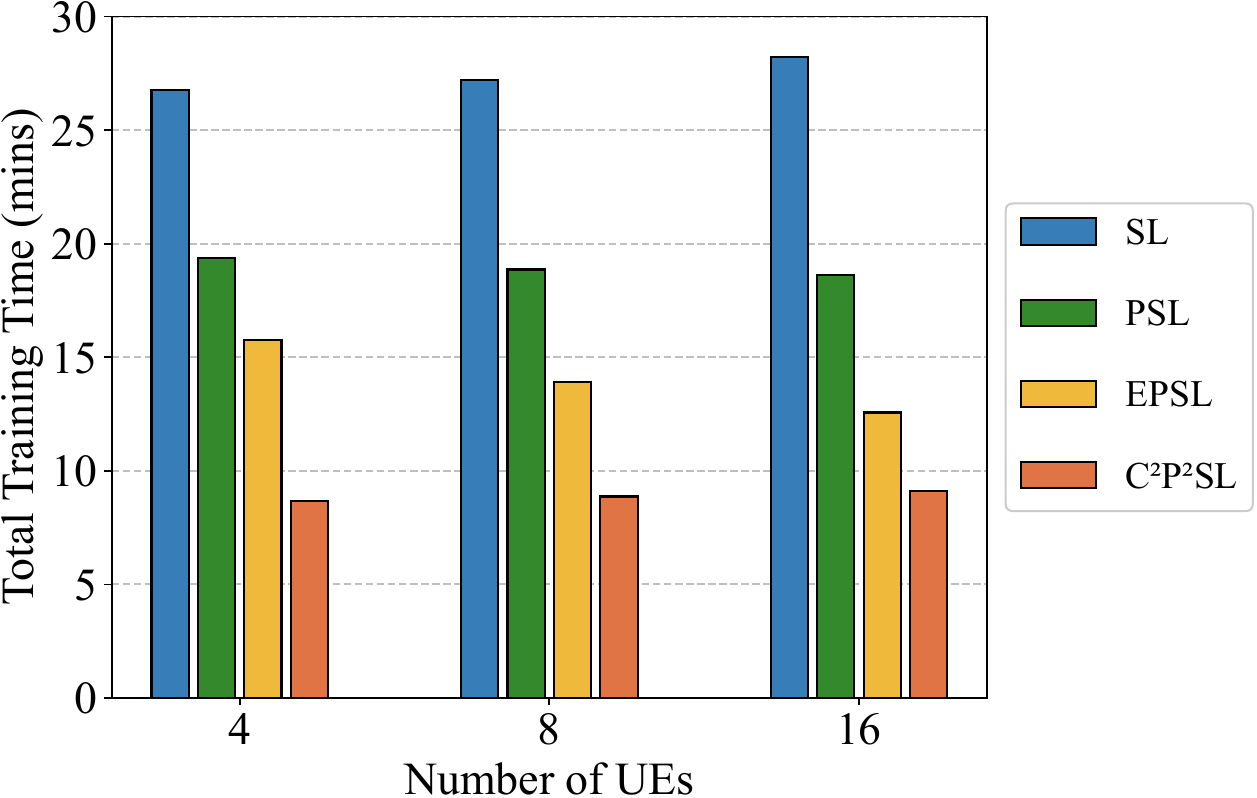}
    \vspace{-0.1cm}
    \caption{Training convergence time of various schemes under different numbers of UEs.}
    \vspace{-0.25cm}
    \label{time}
\end{figure}

Fig.~\ref{time} compares the model convergence time of various schemes under different numbers of UEs to evaluate the wide applications of C$^2$P$^2$SL. The results clearly show that our proposed scheme achieves significantly shorter convergence time than the baseline methods, making an average reduction of 53\% compared to the widely used PSL scheme. This substantially improves training efficiency. It should be noted that our approach maintains superior performance even compared to the EPSL scheme, which accelerates training at the expense of convergence accuracy. Furthermore, since the total dataset size is fixed, the convergence time of each scheme remains almost the same across different UE numbers. This consistency confirms the stability of our scheme's performance improvement under varying scales.

\vspace{-0.25cm}
\begin{figure}[ht]
    \centering
    \hspace{0.6cm}
    \includegraphics[width=0.36\textwidth]{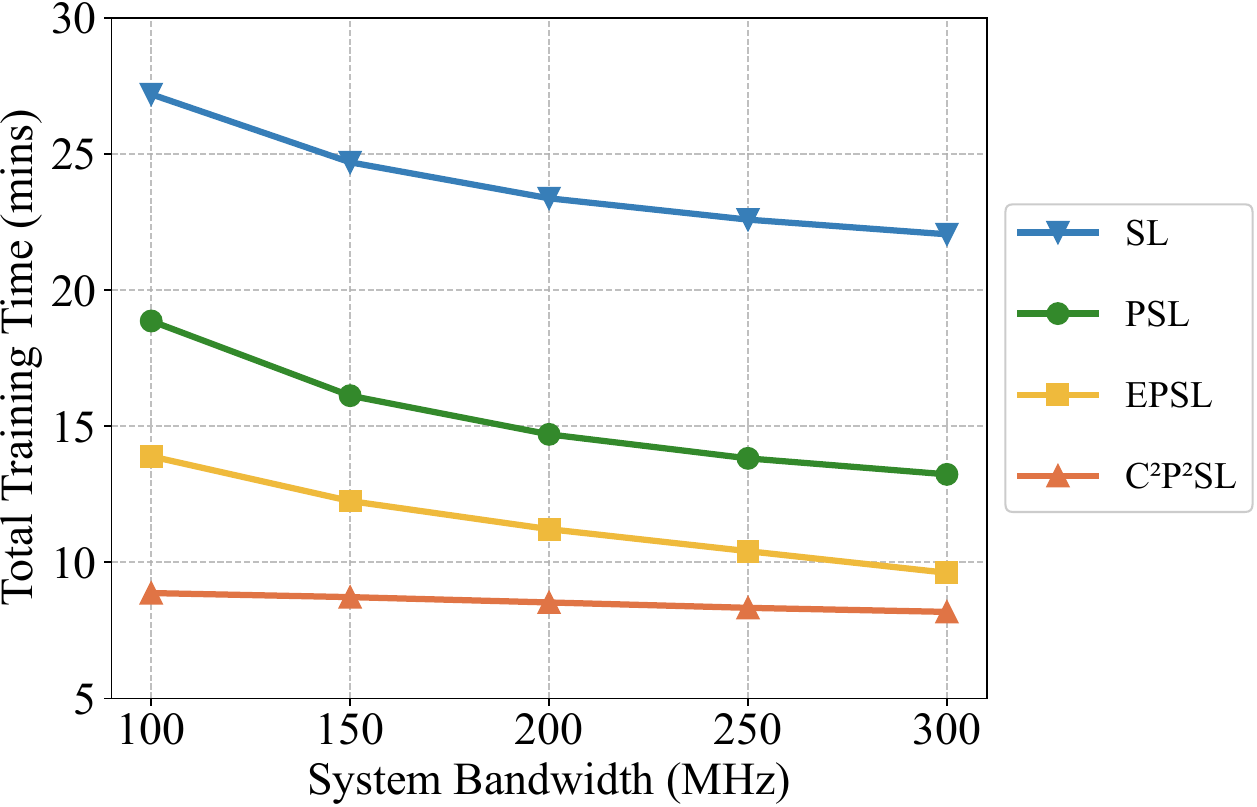}
    \vspace{-0.1cm}
    \caption{Training convergence time of various schemes versus the system bandwidth with n=8.}
    \vspace{-0.25cm}
    \label{bandwidth}
\end{figure}

To further validate the robustness of our proposed scheme, we emulate dynamic wireless channel conditions by varying the total system bandwidth to achieve different uplink and downlink rates. Fig.~\ref{bandwidth} compares the training convergence times of various schemes under varying system bandwidth. The results indicate that our scheme maintains over 38\% reduction against PSL in convergence time across system bandwidths from 100 to 300MHz, showing excellent robustness against different communication conditions. Our scheme achieves more substantial gains in poorer communication environment. This demonstrates the effectiveness of our joint optimization of batch size and time slot allocation, which significantly reduces the synchronization latency caused by the heterogeneous capabilities of UEs. 

\section{Conclusion} \label{conclusion}
This paper proposes the C$^2$P$^2$SL framework, achieving PP scheduling of computation and communication tasks across multiple micro-batches by splitting the training batch. Considering that training time is affected by both model partitioning and heterogeneity of UEs, and that the parallelism depends on micro-batch number, we designed a splitting and allocation scheme through the joint optimization of these variables. Experimental results demonstrate that our scheme significantly reduces training time while maintaining robustness under varying bandwidth conditions.

\section*{Acknowledgment}
This work was supported in part by National Natural Science Foundation of China under Grant 62394292, in part by Zhejiang Provincial Key Research and Development Program under Grant 2023C01021, and in part by the Fundamental Research Funds for the Central Universities under Grant 226-2024-00069. 

\ifCLASSOPTIONcaptionsoff
  \newpage
\fi


\bibliographystyle{IEEEtran}
\bibliography{IEEEabrv.bib, myabrv.bib, ref.bib}

\end{document}